%% file: main.tex
\documentclass[proceedings]{IEEEtran}

\ifCLASSOPTIONcompsoc
  \usepackage[nocompress]{cite}
\else
  \usepackage{cite}
\fi

\usepackage{url}
\makeatletter
\g@addto@macro{\UrlBreaks}{\UrlOrds}
\makeatother

\ifCLASSINFOpdf
  \usepackage[pdftex]{graphicx}
  \graphicspath{{../images/}}
  \DeclareGraphicsExtensions{.pdf,.jpeg,.png}
\else
  \usepackage[dvips]{graphicx}
  \graphicspath{{../eps/}}
\fi

\IEEEoverridecommandlockouts

\begin{document}
%
\title{\textbf{\Large An Integrated Platform for Collaborative Data Analytics}\\}%
%
%
%

\author{\IEEEauthorblockN{Sean Oesch\IEEEauthorrefmark{1},
Rob Gillen, Tom Karnowski}
\IEEEauthorblockA{Oak Ridge National Laboratory\\
\IEEEauthorrefmark{1}oeschts@ornl.gov}}

\author{\IEEEauthorblockN{Sean Oesch\IEEEauthorrefmark{1},
Rob Gillen, Tom Karnowski\\}
\IEEEauthorblockA{Oak Ridge National Laboratory\\
\IEEEauthorrefmark{1}oeschts@ornl.gov}
\IEEEcompsocitemizethanks{\IEEEcompsocthanksitem Authors with Oak Ridge National Laboratory, 
Oak Ridge,
TN, 37830.\protect\\
E-mail: oeschts@ornl.gov, gillenre@ornl.gov, karnowskitp@ornl.gov}
\IEEEcompsocitemizethanks{\IEEEcompsocthanksitem Notice: This manuscript has been authored by UT-Battelle, LLC under Contract No. DE-AC05-00OR22725 with the U.S. Department of Energy. The United States Government retains and the publisher, by accepting the article for publication, acknowledges that the United States Government retains a non-exclusive, paid-up, irrevocable, world-wide license to publish or reproduce the published form of this manuscript, or allow others to do so, for United States Government purposes. The Department of Energy will provide public access to these results of federally sponsored research in accordance with the DOE Public Access Plan (http://energy.gov/downloads/doe-public-access-plan).}
}

\IEEEtitleabstractindextext{%
\begin{abstract}
  While collaboration among data scientists is a key to organizational productivity, data analysts face 
  significant barriers to achieving this end, including data sharing, accessing and configuring the required computational 
  environment, and a unified method of sharing knowledge. Each of these barriers to collaboration is related to the fundamental question of knowledge management ``how can organizations use knowledge more effectively?''. In this paper, we 
  consider the problem of knowledge management in collaborative data analytics and present ShareAL, an integrated knowledge management platform, as a solution to that problem. 
  The ShareAL platform consists of three core components:
  a full stack web application, a dashboard 
  for analyzing streaming data and a High Performance Computing (HPC) cluster for 
  performing real time analysis. Prior research has not applied knowledge management to  
  collaborative analytics or developed a platform with the same capabilities as ShareAL. ShareAL overcomes the barriers 
  data scientists face to collaboration by providing intuitive sharing of data and analytics via the web 
  application, a shared computing environment via the HPC cluster and knowledge sharing and collaboration 
  via a real time messaging application. 
\end{abstract}

}

\maketitle
\thispagestyle{empty}
\pagenumbering{arabic}

\IEEEdisplaynontitleabstractindextext

%
\IEEEpeerreviewmaketitle

\input{introduction}
\input{background}
\input{relatedwork}
\input{approach}
\input{platform}

\input{endmatter}

\bibliographystyle{IEEEtran}
\bibliography{bibtemplate_samples}

%




\end{document}

%% file: introduction.tex
\section{Introduction}\label{introduction}

In 2017 Forbes published an article entitled ``What Are Best Practices For Collaboration Between Data Scientists?''~\cite{forbesarticle} stating that data scientists and data science teams have a history of not effectively collaborating within industry. The existing practices for collaboration within organizations have often developed organically and lack the forethought required to allow data scientists to effectively work together, which is a key to both productivity and profitability~\cite{davenport1998successful}. The article makes the following specific suggestions for improving collaboration among data scientists: work from the same data, version your code, use a data pipeline and share a computational environment. 

The problem these data scientists are experiencing is fundamentally a knowledge management (KM) problem. Barclay defines knowledge as ``information combined with experience, context, interpretation and reflection - a high value form of information that is ready to apply to decisions and actions''~\cite{barclay1997knowledge}. Knowledge management answers the question ``how can organizations use knowledge more effectively?''~\cite{davenport1998successful}. Data scientists have knowledge in the form of data, experience and analytic tools that they use for their daily tasks and the challenge they face is a KM challenge. They need to understand how to manage that knowledge to allow for effective collaboration. 

In this paper, we consider the problem of knowledge management in collaborative data analytics and present ShareAL, an integrated knowledge management platform, as a solution to that problem. ShareAL enables effective knowledge management to allow data analysts to collaborate in the following ways:

\begin{itemize}
    \item intuitive sharing of data and analytics via a web application
    \item a shared computing environment via an HPC cluster
    \item knowledge sharing and collaboration via a real time messaging application
    \item interaction with streaming data via an interactive dashboard
    \item data persistence through effective database management and logging
\end{itemize}

The ShareAL platform consists of three core components: a full stack web application, a dashboard for analyzing streaming data and a High Performance Computing (HPC) cluster for performing real time analysis. Alongside these core components, a local chat server is set up to further facilitate collaboration. ShareAL allows users to create dashboards for each facility of interest, score them via custom analytic products and to share the results of analyses performed on a local HPC Slurm cluster. It enables the sharing of both data and computational environment. In order to allow users to test the platform before investing in HPC related hardware, a virtual HPC cluster implemented in Docker~\cite{docker} is provided for testing purposes. The platform can be accessed from the open source repository~\cite{shareanalytics} of Oak Ridge National Laboratory (ORNL). 

The context for the development of ShareAL was Non-intrusive Load Monitoring (NILM), in which the energy consumption of individual devices within a building is monitored. This information can then be used to improve energy efficiency or, as in our particular case, the determine what types of activities are occurring within a particular building. This process is called facility characterization. For an example of how ShareAL could be used in the real world, please see Section~\ref{example}.

ShareAL differentiates itself from existing analytic platforms by providing out of the box linkage to HPC capabilities, support for facility characterization and real time messaging. In the field of knowledge management, the BioKnow Platform \cite{cassivi2009developing}, while different in design and function than ShareAL, was the most similar KM platform identified in existing literature. Unlike ShareAL, it has no analytic or facility characterization component, has no shared computing environment and utilizes a different set of underlying technologies. Furthermore, prior research has not applied knowledge management principles in collaborative analytics. 

The rest of this paper is laid out as follows. Section~\ref{background} covers the necessary background on knowledge management and facility characterization. Section~\ref{relatedwork} discusses what sets ShareAL apart from existing approaches. Section~\ref{problemandapproach} discusses the theoretical approach used to achieve knowledge management in collaborative data analytics. Section~\ref{platform} provides details on how this approach was implemented. Finally, Section~\ref{conclusion} provides a summary of what was achieved and a description of future work. 

%% file: background.tex
\section{Background}\label{background}

\subsection{Knowledge Management}\label{kmbackground}

According to Davenport~\cite{davenport1998successful}, knowledge is ``information combined with experience, context, interpretation and reflection - a high value form of information that is ready to apply to decisions and actions''. Barclay likewise agrees that knowledge is more than information and requires the ability to interpret and contextualize information~\cite{barclay1997knowledge}. In data analytics, raw data and analytic tools are one form of information, but knowledge entails the ability to understand, interpret and synthesize data, analytic tools and the results of analyses. 

Another way of defining knowledge involves dividing it into two separate categories: explicit knowledge and tacit knowledge~\cite{fengjie2004knowledge}~\cite{nonaka1994dynamic}. Explicit knowledge can be communicated easily through writing, whereas tacit knowledge is difficult to convert into writing and is highly contextualized. Barclay~\cite{barclay1997knowledge} adds that tacit knowledge is influenced by personal opinions or beliefs and therefore requires social context to understand. 

For the remainder of this paper, knowledge will be defined as consisting of explicit and tacit components. Explicit knowledge will consist of information and data that requires minimal context and can be effectively communicated by one individual. Tacit knowledge will consist of knowledge that requires additional context or experience and often engagement through some form of conversation, narrative or experience to comprehend.  

The goal of knowledge management, or KM, is to answer the question ``how can organizations use knowledge more effectively?''~\cite{davenport1998successful}. This question is crucial to the success of any organization or enterprise, since effective KM processes are correlated to a company's ability to generate intellectual capital and influence its capacity to innovate~\cite{cassivi2009developing}. 

There are two fundamental types of platforms for achieving knowledge management - knowledge platforms and social platforms. In knowledge platforms, ``knowledge is stocked in structured directories in order to facilitate the usage of the information''~\cite{cassivi2009developing}. According to Nonaka~\cite{nonaka1994dynamic}
, socialization platforms help users identify sources of tacit knowledge for the sake of transferring that knowledge. In that sense, these platforms recognize that ``sensemaking is often a social process involving parallelization of effort''~\cite{heer2008design}. In essence, knowledge platforms are more concerned with explicit knowledge or information and socialization platforms seek to allow the transfer of tacit or contextual knowledge. An integrated platform contains elements of both a knowledge and a socialization platform. 



\subsection{Facility Characterization via NILM}\label{nilmbackground}

In Non-intrusive Load Monitoring (NILM) a Non-intrusive Appliance Load Monitor (NALM) can monitor power usage of multiple devices on a single circuit without having to be directly connected to each device. The NALM performs an ``analysis of the current and voltage waveforms of the total load and estimates the number and nature of the individual loads, their individual energy consumption, and other relevant statistics such as time-of-day variations''~\cite{hart1992nonintrusive}. Previous work has investigated using NILM to enhance electricity audits of commercial buildings~\cite{berges2010enhancing} and to understand domestic energy consumption~\cite{costanza2012understanding}. 

One important application space for NILM technologies is green energy. By monitoring energy usage patterns and linking those patterns to occupant behavior, changes in lifestyle can be assumed in order to reduce cost or to adjust to the current level of energy available from green energy sources. In fact, systems such as those proposed by Lin~\cite{lin2015advanced} use NILM technologies to adjust energy usage with zero intervention from users. 

Law enforcement has used raw energy usage to identify clandestine methamphetamine laboratories for some time. They reason that a ``seemingly vacant residence with a large, continuous demand for energy may indicate a meth lab''~\cite{clandestinemethlab}. NILM offers a way of analyzing energy usage in a more refined manner by both identifying what types of devices are being run and understanding how much energy each device consumes using NALMs. These energy signatures could then be used to identify whether a building is strictly residential or whether it is being used for other purposes. Analysis can be achieved with minimal intrusion. 

%% file: relatedwork.tex
\section{Related Work}\label{relatedwork}

Prior research has not considered the application of knowledge management practices to collaborative analytics. Because ShareAL is both an integrated knowledge management platform and an analytics platform, it is necessary to  differentiate it from existing implementations in each category. It sets itself apart from the former by demonstrating a unique combination of capabilities for facility characterization. It sets itself apart from the latter because no known prior work considers the challenges of knowledge management in collaborative analytics for facility characterization. 

There are a myriad of existing streaming analytic platforms~\cite{adafruit}~\cite{thinger}~\cite{ubidots}~\cite{thingsboard}~\cite{ibmcloud}~\cite{iotaws} that focus on optimizing the pipeline for converting streaming data to actionable analytics. ShareAL differentiates itself by providing out of the box linkage to HPC capabilities and support for NILM facility characterization. These functionalities are critical for collaborative analytics for facility characterization. 

In the field of knowledge management, the BioKnow Platform \cite{cassivi2009developing}, while different in design and function than ShareAL, was the most similar KM platform identified in existing literature. BioKnow was designed for a consortium of partners involved in bioindustry research and focused on strategic monitoring and intelligence, academic monitoring and intelligence, creation of an expert knowledge base and the development of a bioindustry firm directory. Like ShareAL, it stores information in databases and attempts to share both explicit and tacit knowledge. Unlike ShareAL, it has no analytic or facility characterization component, has no shared computing environment and utilizes a different set of underlying technologies. 

Both the Genome Modeling System (GMS) \cite{griffith2015genome} and DisGeNET~\cite{pinero2020disgenet} are KM platforms developed for genomics. And Lemaignan developed the Open Robots Ontology (ORO) for linking symbols to concepts in the field of robotics \cite{lemaignan2010oro}. These are good examples of KM platforms, but are different in the way they represent data and interface with the user than ShareAL because functional requirements and implementation are closely tied to context. While general purpose KM platforms are used within corporations for content curation and employee engagement, they lack the analytic tools and data ingestion capabilities necessary in collaborative analytics for facility characterization. 

%% file: approach.tex
\section{Problem and Approach}\label{problemandapproach}

In this section the problem of achieving efficient knowledge sharing and effective knowledge management in collaborative data analytics is defined and our approach towards the problem presented. In short, analysts face unique barriers to knowledge sharing because they need to be able to share analytics and data and an efficient means of performing exploratory analysis. Without a shared system, there is a cost both in terms of time and resources to transfer data and analytic tools. Analysts also need a means of engaging in and viewing multi-user discussions regarding the analytic tools to achieve tacit knowledge sharing. The approach to this problem therefore requires an integrated platform and not simply a knowledge platform.

\subsection{The Problem}\label{problem}

We set out to design and build a knowledge management platform for facility characterization analysts. There are already platforms, some open source, that provide analysts with tools for viewing and manipulating streaming data or storing and sharing data sets. However, none of these existing platforms provide a shared computing environment, address the needs of facility characterization in particular or integrate tacit knowledge sharing as a core component. The problem, then, is that an integrated platform that allows multi-user discussions, facilitates the sharing of both analytics and data, provides a shared computing environment and enables tacit knowledge sharing is needed.

\subsection{The Approach}\label{approach}

Our approach to this problem involved building an integrated platform consisting of a web application as the main user interface, an HPC cluster that provides a shared computing environment for exploratory analysis, a team chat application for tacit knowledge sharing and a live dashboard tool for dynamic data engagement. This platform allows the sharing of explicit and tacit knowledge via a user interface designed to reduce knowledge barriers. This platform is a first step towards gaining a deeper understanding of how to achieve effective knowledge management in collaborative analytics. 

\subsubsection{Overview}\label{approachoverview}

\begin{figure}[!t]
    \centering
    \includegraphics[width=\linewidth]{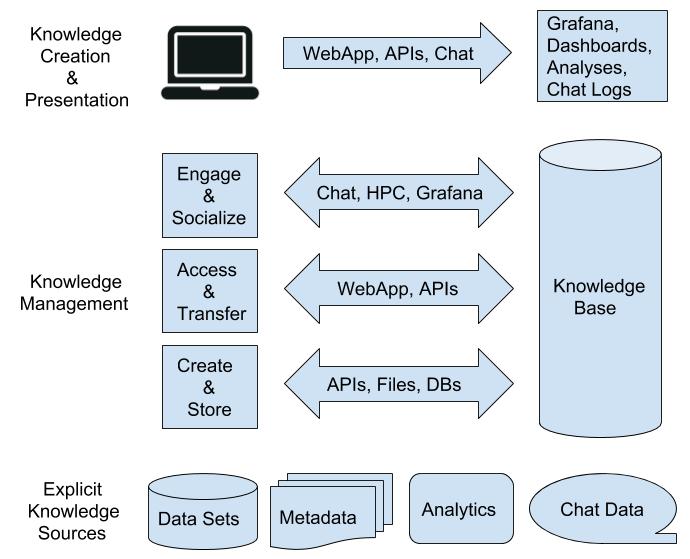}
    \caption{Approach Architecture - an integrated platform consisting of a web application as the main user interface, an HPC cluster that provides a shared computing environment for exploratory analysis, a stand-alone team chat application for tacit knowledge sharing and a live dashboard tool for dynamic data engagement}
    \label{architecture}
\end{figure}

Figure~\ref{architecture} shows a diagram of the architecture of our approach to building an integrated platform for collaborative analytics in facility characterization. This diagram identifies knowledge sources and explains how our approach enables knowledge management and knowledge creation and presentation. Each layer of this diagram represents an ordered stage of the design process. First, all knowledge sources must be identified. Next, a knowledge management strategy must be developed. And finally, to interface the solution with actual users a means of knowledge creation and sharing must be established.  

Explicit knowledge sources include data sets, metadata, analytics and chat data. Data sets can vary in format and the platform must be designed to handle any text based format, such as xml, csv or json, as well as binary data. Data also comes in the form of streaming data from sensors or Internet of Things (IoT) devices. Metadata is data entered by the user on the web application describing when data was collected, how it was collected, when it expires and other information along the same lines. It also includes user information and metadata for analytic tools. Analytics are either executable files or code that can be used to analyze data and deployed on High Performance Computing (HPC) clusters to allow users to perform quick exploratory analysis to determine if they should invest additional time in a particular analytic or data set (this process is also known as data snacking). Chat data is composed of logs from communications between users and can provide valuable contextual information to remove knowledge barriers and share tacit knowledge. 

Knowledge management is achieved in three basic steps: create and store, access and transfer and engage and socialize. Create and store is achieved using APIs for the collection of data and data entered by users from existing databases or files. Access and transfer of knowledge is achieved through the web application and via APIs that can access databases where information is stored. Engage and socialize is the stage that allows the sharing of tacit knowledge and involves user interactions in a team chat application, exploratory analysis achieved through deployment of analytics in an HPC environment and interacting with data via the streaming data a live dashboard. For facility characterization, the web application will provide custom dashboard pages that will allow users to display the status of a particular location along with images of that location and the cumulative score of analytics against relevant data sets. These dashboards will also serve as a means for engaging and socializing knowledge. 

Knowledge creation is achieved via the web application and team chat software running alongside the main web application. This process results in knowledge presented in the form of interactive dashboards, results from analytics run against data sets and chat logs. We will now take a look in more depth at how explicit knowledge and tacit knowledge are stored and transferred and at the functional requirements of the system. 

\subsubsection{Explicit Knowledge Sharing}\label{explicitknowledge}

The first task in allowing explicit knowledge sharing is to consider what types of explicit knowledge exists within collaborative analytics for facility characterization. We identified the following categories:

\begin{itemize}
	\item data sets (static and streaming)
    \item analytics
    \item user information
    \item facility details and status
\end{itemize}

To achieve knowledge sharing of these types of information, only three fundamental elements are needed. A front-end user interface that can be used to input/view the data, a back-end for storing the data and a pipeline for ingesting streaming data. Our design uses two separate databases for static vs streaming data as well as providing separate visualization tools for each. We believe this separation of concerns is optimal both in terms of design and function. 

\subsubsection{Tacit Knowledge Sharing}\label{tacitknowledge}

The first task in allowing tacit knowledge sharing is to consider what types of tacit knowledge exists within collaborative analytics for facility characterization. We identified the following broad categories:

\begin{itemize}
    \item interpreting analytic output
    \item exploratory data analysis and data snacking
    \item understanding when to use an analytic
    \item understanding why and how data was collected
    \item understanding facility characteristics and purpose
\end{itemize}

To achieve tacit knowledge sharing, we need two fundamental elements. First, a means for users to run analytics provided by other users in order to perform exploratory analysis. And second, a way for users to socialize and overcome knowledge barriers via communal engagement. Our approach was to solve the first hurdle within our platform via a local HPC environment where users can run analytic tools against existing data sets. The need for communal engagement is addressed via Let's Chat~\cite{letschat}, an open source chat platform. Any real-time, interactive chat platform (Slack, Mattermost, etc.) could provide similar functionality. A stand-alone Let's Chat server will run alongside the web platform. At a higher level, the entire platform will itself provide a context within which users can participate in tacit knowledge sharing as they interact with the available tools in a team environment. 

\subsubsection{Functional Requirements}\label{requirements}

A software platform has many functional requirements, but in the case of an integrated platform for collaborative analytics in facility characterization there are a few that are significant. First, it is important that the user be able to intuitively create, update and search both data sets and analytics. Second, the user must be able to intuitively add and score new facilities. Third, it is crucial for the conveyance of tacit knowledge that the ability to gain experiential knowledge via running other user's analytics and engaging with the results be accessible. Fourth, the user should be able to multitask and switch between the various tools available in the platform, including team chat.

The method we used to achieve each of these goals involved placing these functionalities in prominent locations in the user interface. All core functions were placed such that they are discoverable when first encountering the user interface and remain discoverable while using the program. Utilizing a web application aids in multitasking, because the concept of tabs helps the user multitask and maintain access to the main website even when switching between tools. Team chat software runs alongside the main application and  can be minimized for easy access while using the platform. 

%% file: platform.tex
\section{The ShareAnalytics Platform}\label{platform}

This section provides details on how we implemented our approach, explained in Section~\ref{approach}, to an integrated collaborative analytics platform for facility characterization. To experiment with the platform, please visit ORNL's public GitHub repository~\cite{shareanalytics}, where full documentation is provided for how to get started. The ShareAnalytics Platform, or ShareAL, is a lightweight and extensible knowledge management platform for collaborative data analytics that meets the needs of NILM facility characterization. It consists of three core components: a full stack web application based on the MEANJS project~\cite{meanjs}, a Grafana~\cite{grafana} dashboard for analyzing streaming data and a High Performance Computing (HPC) cluster running Slurm~\cite{slurm} to facilitate exploratory analysis. ShareAL allows users to create dashboards for each facility of interest, score them via custom analytic products and to share the results of analyses performed on a local HPC Slurm cluster. It enables the sharing of both data and a computational environment for exploratory analysis. A demonstration HPC environment implemented using Docker~\cite{docker} is provided to allow users to test the platform before investing in HPC related hardware if their use case requires additional computing resources.

\begin{figure}[!t]
    \centering
    \includegraphics[width=\linewidth]{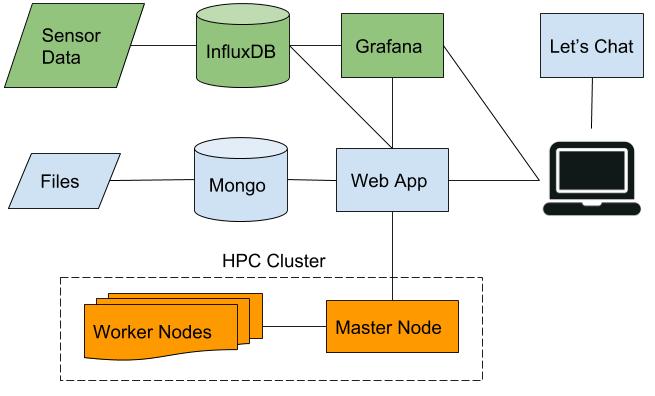}
    \caption{Implementation Architecture - a lightweight and extensible knowledge management platform for collaborative data analytics that meets the needs of NILM facility characterization consisting of a full stack web application, a dashboard for analyzing streaming data and a High Performance Computing (HPC) cluster to facilitate exploratory analysis}
    \label{implementation}
\end{figure}

Figure~\ref{implementation} shows how the core components of ShareAL are connected and how the user interfaces with those components. The web application is the main method of interfacing with the platform. Streaming data can be viewed in Grafana or accessed via the main web application to generate static data sets by providing a date range. Running analytics on the HPC cluster is achieved directly through the web application. Let's Chat~\cite{letschat}~\cite{letschatdockerhub} is open source team chat software that runs alongside the web interface as a stand-alone application. 

\subsection{Web Application}

The web application targets NILM facility characterization. In addition to normal website functionality such as user management, it gives the user access to dashboards, data sets, analytics and the ability to perform analyses on the HPC cluster. The dashboard is geared towards facility characterization and allows the user to create a metric composed of multiple analytics for a particular site to provide a cumulative score. Data sets and analytics can be shared amongst users to allow collaboration or access to the data can be restricted. Once analytics and data sets have been uploaded to the website, an analysis can then be run on the HPC cluster and the results applied to a particular dashboard's facility characterization score or viewed on the results page.

\begin{figure}[!t]
    \centering
    \includegraphics[width=\linewidth]{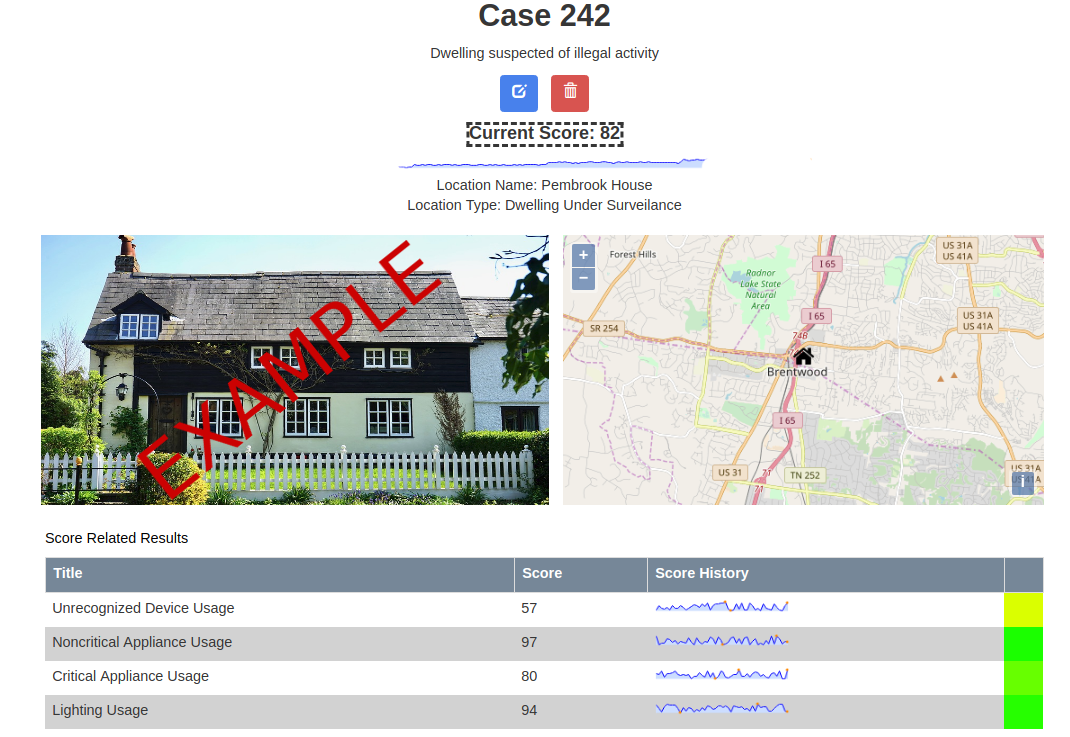}
    \caption{Fictitious Dashboard Example - a simulated site under surveillance by law enforcement using NILM to view historical energy usage and determine the probability that the facility is being used for illegal activity by scoring custom analytics}
    \label{webappimage}
\end{figure}

Figure~\ref{webappimage} shows a fictitious dashboard example for a site under surveillance by law enforcement using NILM. It provides both the historical and current scores of each analytic that contributes to the total score as well as for the total score itself, in this case the score being related to potential illegal activity. Each of the analytics is related to energy usage of a particular category of devices and the total score is a combination of those individual metrics.

\subsection{Dashboard}

\begin{figure}[!t]
    \centering
    \includegraphics[width=\linewidth]{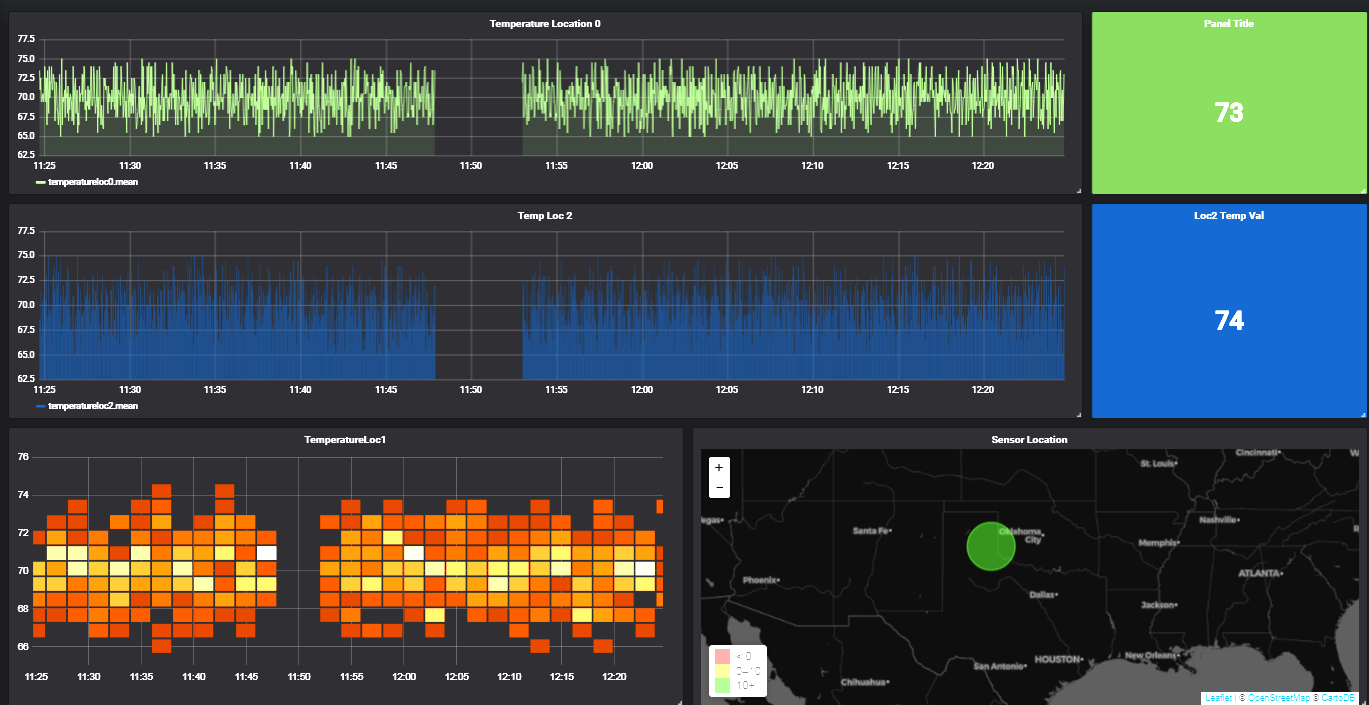}
    \caption{Dashboard Snapshot - a sample grafana dashboard showcasing a small sample of the customizable widgets that can be used to interact with data - graphs, gauges, heat maps and topological maps}
    \label{dashboardimage}
\end{figure}

The dashboard, shown in Figure~\ref{dashboardimage}, is built using Grafana and InfluxDB. Grafana is an interactive dashboard tool that includes drag and drop widgets that reduce barriers to engaging with streaming data. In addition, there are a multitude of plugins that provide additional widgets for specific types of graphs or figures, reducing the amount of work for the user. Instructions for adding plugins are available in the documentation. In addition, Grafana itself is open source. 

\subsection{HPC Cluster}

The HPC cluster allows C, python, Matlab and bash script based analytics to be run against uploaded data sets. Additional languages can be supported by simply editing the web server on the head node of the cluster that handles requests from the web based application. This feature allows users to test the full functionality of the platform without any investment in additional hardware and can be used in deployment if it is sufficient for their use case.

\subsection{Example Use Case}\label{example}

A power company wants to use NILM to monitor energy usage in a specific set of communities so that they can better understand their customers. Their statistician has developed a set of analytics to achieve this task and uploaded them to ShareAL via the web interface. At the same time, an analyst from law enforcement has written a script to determine if a residence is being used for illegal activities, uploaded it to ShareAL and asked the power company to run it as they ingest new data (with relevant legal permissions in place and in accordance with privacy laws). When the power company's analyst checks their dashboards, they see an alert that there may be illegal activity and contact the law enforcement analyst through ShareAL's chat application. Via knowledge gained in this conversation, the law enforcement analyst decides to log in to ShareAL and run some additional analytics. ShareAL allowed users from different organizations to share and generate knowledge by removing common barriers to communication and data analysis. Note that this is a conceptual example and assumes proper warrants and legal permissions are in place as described.

%% file: endmatter.tex
\section{Conclusion and Future Work}\label{conclusion}

In this paper we considered the problem of knowledge management in collaborative data analytics and presented ShareAL, an integrated knowledge management platform, as a solution to that problem for facility characterization. Data science presents unique barriers to effective collaboration. At a minimum, data scientists must have the ability to share both data and a computational environment to collaborate effectively~\cite{forbesarticle}. In addition to these basic properties, the solution we have presented allows the sharing of both explicit and tacit knowledge among scientists. ShareAL accomplishes these goals via a web application with a shared compute environment, an interactive dashboard tool and Let's Chat for team chat. 
One thread of future work involves deploying the ShareAL platform in varied environments to see how well it removes the perceived barriers to knowledge sharing and to determine what barriers still remain. While ShareAL was designed for facility characterization, it can be adapted to other environments to address the challenges of collaborative analytics. In the process of making these adaptations, the platform can be extended and improved. 

Future work in the theory of knowledge management for data analytics also needs to be done. Collaborative analytics provides a unique context for applying knowledge management theory and practices. There is still plenty of room for growth in defining terminology within this specific context and refining the model for achieving knowledge sharing, removing knowledge barriers and ensuring that both explicit and tacit knowledge are transferred effectively. 

%% file: main.bbl
\begin{thebibliography}{10}
\providecommand{\url}[1]{#1}
\csname url@samestyle\endcsname
\providecommand{\newblock}{\relax}
\providecommand{\bibinfo}[2]{#2}
\providecommand{\BIBentrySTDinterwordspacing}{\spaceskip=0pt\relax}
\providecommand{\BIBentryALTinterwordstretchfactor}{4}
\providecommand{\BIBentryALTinterwordspacing}{\spaceskip=\fontdimen2\font plus
\BIBentryALTinterwordstretchfactor\fontdimen3\font minus
  \fontdimen4\font\relax}
\providecommand{\BIBforeignlanguage}[2]{{%
\expandafter\ifx\csname l@#1\endcsname\relax
\typeout{** WARNING: IEEEtran.bst: No hyphenation pattern has been}%
\typeout{** loaded for the language `#1'. Using the pattern for}%
\typeout{** the default language instead.}%
\else
\language=\csname l@#1\endcsname
\fi
#2}}
\providecommand{\BIBdecl}{\relax}
\BIBdecl

\bibitem{forbesarticle}
\BIBentryALTinterwordspacing
B.~Hammer. (2017, April) What are best practices for collaboration between data
  scientists? [Online]. Available:
  \url{https://www.forbes.com/sites/quora/2017/04/04/what-are-best-practices-for-collaboration-between-data-scientists/}
\BIBentrySTDinterwordspacing

\bibitem{davenport1998successful}
T.~H. Davenport, D.~W. De~Long, and M.~C. Beers, ``Successful knowledge
  management projects,'' \emph{Sloan management review}, vol.~39, no.~2, pp.
  43--57, 1998.

\bibitem{barclay1997knowledge}
R.~O. Barclay and P.~C. Murray, ``What is knowledge management,''
  \emph{Knowledge praxis}, vol.~19, 1997.

\bibitem{docker}
\BIBentryALTinterwordspacing
docker. [Online]. Available: \url{https://www.docker.com/}
\BIBentrySTDinterwordspacing

\bibitem{shareanalytics}
\BIBentryALTinterwordspacing
Shareanalytics {Open} {Source} {Repository}. [Online]. Available:
  \url{https://github.com/ORNL/ShareAnalytics}
\BIBentrySTDinterwordspacing

\bibitem{cassivi2009developing}
L.~Cassivi, A.-L. Saives, E.~Labzagui, and P.~Hadaya, ``Developing a knowledge
  sharing platform: the case of a bio-industry research consortium,'' in
  \emph{Information, Process, and Knowledge Management, 2009. eKNOW'09.
  International Conference on}.\hskip 1em plus 0.5em minus 0.4em\relax IEEE,
  2009, pp. 153--158.

\bibitem{fengjie2004knowledge}
A.~Fengjie, Q.~Fei, and C.~Xin, ``Knowledge sharing and web-based
  knowledge-sharing platform,'' in \emph{E-Commerce Technology for Dynamic
  E-Business, 2004. IEEE International Conference on}.\hskip 1em plus 0.5em
  minus 0.4em\relax IEEE, 2004, pp. 278--281.

\bibitem{nonaka1994dynamic}
I.~Nonaka, ``A dynamic theory of organizational knowledge creation,''
  \emph{Organization science}, vol.~5, no.~1, pp. 14--37, 1994.

\bibitem{heer2008design}
J.~Heer and M.~Agrawala, ``Design considerations for collaborative visual
  analytics,'' \emph{Information visualization}, vol.~7, no.~1, pp. 49--62,
  2008.

\bibitem{hart1992nonintrusive}
G.~W. Hart, ``Nonintrusive appliance load monitoring,'' \emph{Proceedings of
  the IEEE}, vol.~80, no.~12, pp. 1870--1891, 1992.

\bibitem{berges2010enhancing}
M.~E. Berges, E.~Goldman, H.~S. Matthews, and L.~Soibelman, ``Enhancing
  electricity audits in residential buildings with nonintrusive load
  monitoring,'' \emph{Journal of industrial ecology}, vol.~14, no.~5, pp.
  844--858, 2010.

\bibitem{costanza2012understanding}
E.~Costanza, S.~D. Ramchurn, and N.~R. Jennings, ``Understanding domestic
  energy consumption through interactive visualisation: a field study,'' in
  \emph{Proceedings of the 2012 ACM Conference on Ubiquitous Computing}.\hskip
  1em plus 0.5em minus 0.4em\relax ACM, 2012, pp. 216--225.

\bibitem{lin2015advanced}
Y.-H. Lin and M.-S. Tsai, ``An advanced home energy management system
  facilitated by nonintrusive load monitoring with automated multiobjective
  power scheduling,'' \emph{IEEE Transactions on Smart Grid}, vol.~6, no.~4,
  pp. 1839--1851, 2015.

\bibitem{clandestinemethlab}
\BIBentryALTinterwordspacing
J.~Petrocelli. Clandestine {M}eth {L}abs. [Online]. Available:
  \url{https://www.policemag.com/channel/patrol/articles/2009/01/clandestine-meth-labs/page/2.aspx}
\BIBentrySTDinterwordspacing

\bibitem{adafruit}
\BIBentryALTinterwordspacing
io.adafruit. [Online]. Available: \url{https://io.adafruit.com/}
\BIBentrySTDinterwordspacing

\bibitem{thinger}
\BIBentryALTinterwordspacing
thinger.io. [Online]. Available: \url{https://thinger.io/}
\BIBentrySTDinterwordspacing

\bibitem{ubidots}
\BIBentryALTinterwordspacing
ubidots. [Online]. Available: \url{https://ubidots.com/}
\BIBentrySTDinterwordspacing

\bibitem{thingsboard}
\BIBentryALTinterwordspacing
thingsboard.io. [Online]. Available: \url{https://thingsboard.io/}
\BIBentrySTDinterwordspacing

\bibitem{ibmcloud}
\BIBentryALTinterwordspacing
{IBM} {IoT} analytics on the cloud. [Online]. Available:
  \url{https://internetofthings.ibmcloud.com/}
\BIBentrySTDinterwordspacing

\bibitem{iotaws}
\BIBentryALTinterwordspacing
iotaws. [Online]. Available: \url{https://aws.amazon.com/iot/}
\BIBentrySTDinterwordspacing

\bibitem{griffith2015genome}
M.~Griffith, O.~L. Griffith, S.~M. Smith, A.~Ramu, M.~B. Callaway, A.~M.
  Brummett, M.~J. Kiwala, A.~C. Coffman, A.~A. Regier, B.~J. Oberkfell
  \emph{et~al.}, ``Genome modeling system: a knowledge management platform for
  genomics,'' \emph{PLoS computational biology}, vol.~11, no.~7, p. e1004274,
  2015.

\bibitem{pinero2020disgenet}
J.~Pi{\~n}ero, J.~M. Ram{\'\i}rez-Anguita, J.~Sa{\"u}ch-Pitarch, F.~Ronzano,
  E.~Centeno, F.~Sanz, and L.~I. Furlong, ``The disgenet knowledge platform for
  disease genomics: 2019 update,'' \emph{Nucleic acids research}, vol.~48,
  no.~D1, pp. D845--D855, 2020.

\bibitem{lemaignan2010oro}
S.~Lemaignan, R.~Ros, L.~M{\"o}senlechner, R.~Alami, and M.~Beetz, ``Oro, a
  knowledge management platform for cognitive architectures in robotics.'' in
  \emph{IROS}, 2010, pp. 3548--3553.

\bibitem{letschat}
\BIBentryALTinterwordspacing
Let's chat. [Online]. Available: \url{http://sdelements.github.io/lets-chat/}
\BIBentrySTDinterwordspacing

\bibitem{meanjs}
\BIBentryALTinterwordspacing
meanjs. [Online]. Available: \url{http://meanjs.org/}
\BIBentrySTDinterwordspacing

\bibitem{grafana}
\BIBentryALTinterwordspacing
grafana. [Online]. Available: \url{https://grafana.com/}
\BIBentrySTDinterwordspacing

\bibitem{slurm}
\BIBentryALTinterwordspacing
slurm. [Online]. Available: \url{https://slurm.schedmd.com/}
\BIBentrySTDinterwordspacing

\bibitem{letschatdockerhub}
\BIBentryALTinterwordspacing
Let's {Chat} {Docker} {Hub} repository. [Online]. Available:
  \url{https://hub.docker.com/r/sdelements/lets-chat/}
\BIBentrySTDinterwordspacing

\end{thebibliography}
